# *Driving Forces and Boundary Conditions in Continuum Dislocation Mechanics*

Amit Acharya

Dept. of Civil and Environmental Engineering

Carnegie Mellon University, Pittsburgh, PA 15213, U.S.A.[*]

**Summary**

As a guide to constitutive specification, driving forces for dislocation velocity and nucleation rates are derived for a field theory of dislocation mechanics and crystal plasticity proposed in Acharya (2001, *J. Mech. Phys. Solids*). A condition of closure for the theory in the form of a boundary condition for dislocation density evolution is also derived. The closure condition is generated from a uniqueness analysis in the linear setting for partial differential equations controlling the evolution of dislocation density. The boundary condition has a simple physical meaning as an inward flux over the dislocation inflow part of the boundary. Kinematical features of dislocation evolution like initiation of bowing of a pinned screw segment, and the initiation of cross-slip of a single screw segment are discussed. An exact solution representing the expansion of a polygonal dislocation loop is derived for a quasilinear system of governing partial differential equations. The representation within the theory of some physical features of dislocation mechanics and plastic deformation like local (dislocation level) Schmid and non-Schmid behavior, unloaded, stress free and steady microstructures, and yielding are also discussed.

## *1. Introduction*

This paper aims at advancing the theory of continuum dislocation mechanics presented in Acharya (2001) (referred to as [I] henceforth) with respect to providing general guidelines for the formulation of constitutive equations and the specification of boundary conditions for dislocation density evolution. The associated discussion also considers the representation within the theory of such physical behaviors as expansion of a polygonal dislocation loop and the increase of the norm of dislocation density from a prescribed initial state purely due to the kinematics of dislocation density evolution; kinematics of the initiation of bowing of a pinned screw segment and the (conceptually similar) initiation of cross-slip of a single screw segment; local (dislocation level) Schmid and non-Schmid behavior; unloaded, stress free (zero energy) and steady microstructures; and yielding.

---

[*] Tel. (412) 268 4566; Fax. (412) 268 7813; email: amita@andrew.cmu.edu

The theory developed in this paper, and its predecessor [I], is a *continuum* theory in the sense that it is a mathematical idealization of the stressing and deformation of crystalline materials. It deals with a mathematical continuum of points, endowed with physically motivated attributes, and such a continuum serves as an idealization of a crystalline body. In their conception, its ingredients are not explicitly built on rigorous averages of discrete physical quantities (e.g. dislocation densities) – however, the hope is that its solutions model observable physical quantities related to dislocation mechanics in crystals, and its success, or failure, is to be judged purely by the closeness of such solutions to actual physical behavior. As a result of this interpretation, we shall not be unduly concerned about the philosophical implications of sometimes trying to model individual dislocation behavior with this theory, where an individual dislocation is modeled as a suitable variation of the dislocation density field, e. g. a non-zero distribution in a cylinder along a curve in space. In fact, at every appropriate opportunity we apply this continuum theory of dislocations to the mechanics of single dislocations in order to test its soundness and glean important modeling information from such efforts. To some extent, this program has been carried out in [I], and we use this idea repeatedly in the paper.

In the conception of a well-posed, non-equilibrium continuum theory suitable for the study of stress, work-hardening, and permanent deformation of dislocation distributions, this work is motivated by the works of Kröner (1981), Mura (1963, 1970), Willis (1967), Kosevich (1979), Fox (1966), Head *et al.* (1993), Aifantis (1986), and van der Giessen and Needleman (1995).

A recent thrust in continuum crystal plasticity modeling has been to account for 'geometrically necessary' dislocations, while using ideas from slip-based conventional crystal plasticity to model the physical effects of 'statistically stored' ones. Since the idea of geometrically necessary and statistically stored dislocations is necessarily related to a scale of spatial resolution, it may also be useful to view the problem of dislocation mechanics, and the associated crystal plasticity, at a sufficiently small physical scale (inter-dislocation spacing) so that all dislocations are 'geometrically necessary'. There is a large class of current and emerging technological problems for which such a theory would be directly applicable. Moreover, if a continuum framework for the *nonequilibrium* behavior of dislocation distributions can be developed with such a fine scale of resolution, then an effort can be made to average such a theory (e. g. Muncaster, 1983 a,b) to obtain others with more macroscopic scales of resolution that have naturally built into them the geometric and physical attributes, and effects, of 'statistically stored' as well as 'geometrically necessary' dislocation distributions. The points mentioned above serve as partial motivation for the theory being developed in this paper.

## *2. Notation and some Results from Potential Theory*

The symbol $\in$ is shorthand for 'belongs to'; $\forall$, that for 'for all'; $\subset$ stands for 'subset' and $\Rightarrow$ for 'implies'. $\text{sgn}(a)$ denotes the sign of the argument '$a$'. A superposed dot on a symbol represents a material time derivative. The statement $a := b$ is meant to indicate that $a$ is being defined to be equal to $b$. The summation convention is implied except for indices appearing between parenthesis or when explicitly mentioned to the contrary. We denote by $\boldsymbol{Ab}$ the action of the second-order (third-order, fourth-order) tensor $\boldsymbol{A}$ on the vector (second-order tensor, second-order tensor) $\boldsymbol{b}$, producing a vector (vector, second-order tensor). A $\cdot$ represents the inner product of two vectors, a : represents the trace inner product of two second-order tensors (in rectangular Cartesian components, $\boldsymbol{A}:\boldsymbol{B} = A_{ij}B_{ij}$) and matrices. The symbol $\boldsymbol{AB}$ represents tensor multiplication of the second-order tensors $\boldsymbol{A}$ and $\boldsymbol{B}$. The *curl* operation and the cross product of a second-order tensor and a vector are defined in analogy with the vectorial case and the divergence of a second-order tensor: for a second-order tensor $\boldsymbol{A}$, a vector $\boldsymbol{v}$, and a spatially constant vector field $\boldsymbol{c}$,

$$\begin{aligned}(\boldsymbol{A}\times\boldsymbol{v})^T \boldsymbol{c} &= (\boldsymbol{A}^T\boldsymbol{c})\times\boldsymbol{v} \quad \forall \boldsymbol{c} \\ (curl\,\boldsymbol{A})^T \boldsymbol{c} &= curl(\boldsymbol{A}^T\boldsymbol{c}) \quad \forall \boldsymbol{c}. \end{aligned} \quad (1)$$

In rectangular Cartesian components,

$$\begin{aligned}(\boldsymbol{A}\times\boldsymbol{v})_{im} &= e_{mjk} A_{ij} v_k \\ (curl\,\boldsymbol{A})_{im} &= e_{mjk} A_{ik,j}, \end{aligned} \quad (2)$$

where $e_{mjk}$ is a component of the third-order alternating tensor $\boldsymbol{X}$. The definition for the *curl* of a second order tensor field above follows that of Cermelli and Gurtin (2001) up to a transpose.

The notation $\boldsymbol{A}_{//}$ represents the orthogonal projection[i] of the second-order tensor field $\boldsymbol{A}$ on the null space of the operator *curl* defined in the weak sense [I]. Without loss of generality, in this paper we modify slightly the weak definition of the operator *curl* and its null space. In [I], the problem $curl\, W = -\alpha$ was defined weakly as follows:

$$\int_R W_{rk}\, e_{kji} Q_{ri,j}\, dv = -\int_R \alpha_{ri} Q_{ri}\, dv \quad \text{for all } Q_{ri} \text{ in } T, \tag{3}$$

where all components are with respect to a rectangular Cartesian basis. $T$ is the set of continuous test functions on the domain $R$ with vanishing tangential component on the boundary $\partial R$ of $R$, and at least piece-wise continuous first derivatives in $R$. From hereon, we consider *all test functions belonging to the set $T$ to vanish on the boundary of the body instead of only their tangential components*. With this minor change in the definition of the set $T$, all other definitions and results in [I] remain in force. In particular, the null space, $N(curl)$, of the operator *curl* is defined as the set of all square integrable matrix fields $W$ on $R$ that satisfy

$$\int_R W : curl\, Q\, dv = 0, \quad \forall Q \in T. \tag{4}$$

---

[i] Usually, the terminology 'projection' refers to the projection operator in linear algebra. Here we prefer to call the operator the 'projector', and the result of its action on a member of its domain as the 'projection' of the element being projected.

In other words, $N(curl)$ could be referred to as the set of all weakly irrotational matrix fields.

In this paper we shall also need the following results that are obvious extensions of the work of Weyl (1940) to the matrix case. Let $D$ be the set of all square integrable matrix fields on the domain $R$, endowed with the standard $L_2$ inner product – the inner product of two matrix fields $A$, $B$ is given by $\int_R A : B\, dv$. Then, any $A \in D$ can be *uniquely* written as the sum

$$A = A_\perp + A_{//} \tag{5}$$

where

i. $A_{//}$ belongs to $N(curl) \subset D$.

ii. $A_\perp$ belongs to $N_\perp(curl) = \left\{ B \in D \text{ such that } \int_R B : V\, dv = 0, \forall V \in N(curl) \right\}$. The space $N_\perp(curl)$ is the closure of the set of all tensor fields of the type $curl\, Q^*$ on $R$, $Q^* \in T$, and $Q^*$ sufficiently smooth for $curl\, Q^*$ to make sense.

iii. $\int_R A : B\, dv = \int_R A_{//} : B_{//}\, dv + \int_R A_\perp : B_\perp\, dv$.

iv. For every $A \in D$ there exists a $W_A \in D$ that satisfies

$$\int_R W_A : curl\, Q\, dv = \int_R A_\perp : Q\, dv \quad \forall Q \in T, \text{ with } W_A = 0 \text{ on } \partial R. \tag{6}$$

$W_A$ may be determined by solving

$$\left.\begin{array}{l}\operatorname{curl} W_A = A_\perp \\ \operatorname{div} W_A = 0\end{array}\right\} \text{ on } R \qquad (7)$$
$$W_A = 0 \text{ on } \partial R,$$

or the 'weak' equivalent of (7) when $A$ is not sufficiently smooth.

## 3. Field Equations

For ease of presentation of the main physical ideas without the necessary subtleties that arise from a consideration of finite deformations, this paper will deal only with the theory for small deformations. Most of the ideas presented are, however, generalizable to the finite deformation case, and such a generalization for all of the ideas presented herein will be the focus of subsequent work.

For reasons that will become clear when the boundary condition/uniqueness analysis for the evolution of dislocation density is presented, we associate only one dislocation velocity vector to each slip system dislocation density tensor. This may be viewed as a special case of the type of dislocation velocity description allowed in [I]. With this understanding the field equations of geometrically linear continuum dislocation mechanics are [I]:

$$\operatorname{curl} U^p = -\alpha \qquad (8)$$

$$U^p_{/\!/} = \tilde{U}^p_{/\!/} \qquad (9)$$

$$T = C\left(\varepsilon - \varepsilon^p\right) \quad ; \quad \varepsilon := \frac{1}{2}\left(U + U^T\right) \quad ; \quad \varepsilon^p := \frac{1}{2}\left(U^p + U^{pT}\right) \qquad (10)$$

$$\operatorname{div} T = 0 \qquad (11)$$

$$\dot{\alpha}^{(\kappa)} = -\operatorname{curl}\left(\alpha^{(\kappa)} \times V^{(\kappa)}\right) + s^{(\kappa)} \quad ; \quad \alpha = \sum_\kappa \alpha^{(\kappa)} \qquad (12)$$

$$\dot{\tilde{U}}^p = \sum_\kappa \alpha^{(\kappa)} \times V^{(\kappa)} . \qquad (13)$$

In the above, $U$ is the displacement gradient, $I$ is the second-order identity and $U^e$, $\tilde{U}^p$, and $U^p$ are measures of 'small' elastic, slip, and plastic deformation respectively (second-order tensors), and $C$ is the fourth-order tensor of linear elastic moduli. Also, $T$ is the (symmetric) stress tensor, $\alpha$ and $\alpha^{(\kappa)}$ the total and $\kappa^{th}$ slip system dislocation density tensors respectively, and $V^{(\kappa)}$ the $\kappa^{th}$ slip system dislocation velocity vector. $s^{(\kappa)}$ represents the dislocation source on the $\kappa^{th}$ slip system. The reason for denoting the slip deformation rate as in (13) is based on the kinematics of the slip deformation increment produced by dislocation motion. Indeed, let $b \otimes l$ represent a discrete dislocation dyad with $b$ as the 'true' Burgers vector of the dislocation and $l$ the (positive) line direction of the dislocation in the dislocated lattice (for conventions, see Willis, 1967). If $v$ is the dislocation velocity, then the local shear increment produced in the time interval $\Delta t$, of the material around the dislocation line, is given by $b \otimes (l \times \Delta t\, v)$. Associated with each slip system are three unit vectors that remain fixed in the small deformation idealization: the slip plane normal $i_1^{(\kappa)}$, the slip direction $i_2^{(\kappa)}$, and the other unit vector in the slip plane that forms an orthonormal triad given by $i_3^{(\kappa)} := i_1^{(\kappa)} \times i_2^{(\kappa)}$.

In order to understand why one might succeed in solving the system (8)-(13), we think of the slip system dislocation densities as the state variables of the theory. Given the body and an initial dislocation density state, (8), (10), and (11) are solved first to determine the state of initial stress and the initial condition on $\tilde{U}^p$ [I]. The evolution equations (12)-(13) provide the forcing functions for the solution of $U^p$ from (8)-(9). The solution for the $U^p$ field and (10)-(11) provide the solution for the total displacement. Of course, appropriate boundary conditions are also required to perform the above calculations uniquely.

The utility of (9) is that it renders the weak solution of (8) and (9) unique. The main idea behind the proof of this assertion may be understood readily from an analogy with the matrix case – if a square matrix $A$ is singular then a solution of the matrix equation $Ax = b$ is non-unique up to addition of any vector from the null space of $A$. If the matrix equation is now augmented by the additional requirement that the null space component of the solution $x$ be a specific vector from the null space of $A$ then, of course, the non-uniqueness is eliminated from the solution of the augmented system. Of course, in the matrix case the linear operator in question has a finite-dimensional (and, consequently, complete) linear space as its domain, so speaking of a null space component of an element in the operator's domain through an orthogonal projection does not require any additional concerns; in the case when the domain is an infinite-dimensional function space, completeness of the function space and the operator's null space is not guaranteed, and a weak formulation of the problem is required. In proceeding further with this matrix analogy, the condition for existence of a weak solution to (8) is that the field $\alpha$ be weakly solenoidal (Weyl, 1940) – in the matrix case this is represented by the fact that a solution to $Ax = b$ exists only if the vector $b$ is in the column space of the operator $A$.

### *4. Kinematic Component of the Slip System Source Terms*

Observations of crystal dislocations suggest that dislocation lines lie on slip planes of the crystal. Their Burgers vectors may be in arbitrary directions, with the most easily movable dislocations having their Burgers vector in a slip direction on the slip plane on which the dislocation line lies. The evolution equations for the slip system dislocation densities (12) do not *a priori* guarantee that $\boldsymbol{\alpha}^{(\kappa)}\boldsymbol{i}_1^{(\kappa)} = \boldsymbol{0}$ at all times for physically appropriate choices of the slip system dislocation velocity and the source. This fact motivates a partial characterization of the slip system sources of a kinematic nature; we choose them to be of the form

$$\boldsymbol{s}^{(\kappa)} := \left[\left\{curl\left(\boldsymbol{\alpha}^{(\kappa)} \times \boldsymbol{V}^{(\kappa)}\right)\right\}\boldsymbol{n}^{(\kappa)}\right] \otimes \boldsymbol{n}^{(\kappa)} + \sum_{i=1}^{3}\sum_{j=2}^{3}\left(f_{ij}^{(\kappa)} + S_{ij}^{(\kappa)}\right) \boldsymbol{i}_i^{(\kappa)} \otimes \boldsymbol{i}_j^{(\kappa)}, \tag{14}$$

where $S_{ij}^{(\kappa)}$ represent crystallographic dislocation nucleation rates (to be specified constitutively) and the $f_{ij}^{(\kappa)}$ arise from a redistribution of the 'normal' part of the slip system dislocation density increments. The latter will be specified as follows.

Denote the sum of all the normal slip system dislocation increments as

$$\boldsymbol{\upsilon} := -\sum_{\kappa}\left[\left\{curl\left(\boldsymbol{\alpha}^{(\kappa)} \times \boldsymbol{V}^{(\kappa)}\right)\right\}\boldsymbol{n}^{(\kappa)}\right] \otimes \boldsymbol{n}^{(\kappa)}. \tag{15}$$

Let $K$ be the total number of slip systems in the material. Consider the linear transformation $\mathfrak{J}$ from the vector space $\mathbb{R}^{6K}$ of all $6K$-tuples of real numbers $\left(\delta f_{ij}^{(\kappa)}\right)$ to the space of second-order tensors denoted symbolically by

$$(\delta f) \mapsto \sum_{\kappa}\sum_{i=1}^{3}\sum_{j=2}^{3}\delta f_{ij}^{(\kappa)}\boldsymbol{i}_i^{(\kappa)} \otimes \boldsymbol{i}_j^{(\kappa)} =: \mathfrak{J}\delta f. \tag{16}$$

We assume that the range space of $\mathfrak{J}$ is the whole space of second-order tensors. $\mathbb{R}^{6K}$ is endowed with the standard inner product $(\delta f, \delta g) = \sum_{\kappa}\sum_{i=1}^{3}\sum_{j=2}^{3}\delta f_{ij}^{(\kappa)}\delta g_{ij}^{(\kappa)}$ for all $\delta f$ and $\delta g$, so that it is possible to speak of an orthogonal projection of any $6K$-tuple $\delta f$ on the null space of the operator $\mathfrak{J}$. The $6K$-tuple $(f)$ appearing in (14) is now *uniquely* determined as the solution of

$$\mathfrak{J} f = \boldsymbol{\upsilon} \tag{17}$$

subject to the *additional condition* that the *projection of $(f)$ on the null space of the operator $\mathfrak{J}$ be the null $6K$-tuple*. The requirement of a vanishing null space component of the solution may be interpreted as requiring that all the redistributed 'crystallographic' dislocation density increments be 'geometrically necessary' (Arsenlis and Parks, 1999), which is appropriate since we have in mind a theory whose spatial resolution is required to be adequate to account for phenomena at the scale of inter-dislocation spacing.

The procedure described above may be summarized as a deterministic rule for assigning the forest dislocations arising due to dislocation activity on a system (e.g. cross slip), to other systems whose slip planes can accommodate such dislocations.

A similar mathematical procedure has also been used by Arsenlis and Parks (1999)[ii] in a related but different physical context – our mathematical formulation is also intended to amplify directly the role of all redistributed dislocation density rates being 'geometrically necessary'. There are also obvious similarities in the general idea used here and the one behind the formulation of (9), viewed in the context of delivering uniqueness of solutions to (8).

We view the special case of a material for which the set of dislocation dyads appearing in (16) do not span the space of second-order tensors as a situation where the slip system dislocation velocities have to be constrained so that $\dot{\boldsymbol{\alpha}}^{(\kappa)} \boldsymbol{i}_1^{(\kappa)} = \boldsymbol{0}$. This is an unlikely situation in the case of most crystal structures, since six linearly independent dyads are provided just by one slip system, but it is possible, e.g., for a material with only one slip plane.

## 5. Driving Forces

To identify the driving forces for the slip system dislocation velocity vectors and crystallographic nucleation rates, we examine the *global* (mechanical) dissipation in the theory,

$$D := \int_R \boldsymbol{T} : \dot{\boldsymbol{U}} \, dv - \int_R \dot{\psi} \, dv \tag{18}$$

where $\psi$ is the free energy per unit volume of the body. The free energy, like the stress, is assumed to depend only on the elastic strain, $\boldsymbol{\varepsilon}^e$,

$$\psi = \hat{\psi}\left(\boldsymbol{\varepsilon} - \boldsymbol{\varepsilon}^p\right). \tag{19}$$

Consequently,

$$D = \int_R \left(\boldsymbol{T} - \frac{\partial \hat{\psi}}{\partial \boldsymbol{\varepsilon}^e}\right) : \dot{\boldsymbol{\varepsilon}} \, dv + \int_R \frac{\partial \hat{\psi}}{\partial \boldsymbol{\varepsilon}^e} : \dot{\boldsymbol{\varepsilon}}^p \, dv . \tag{20}$$

We assume hyperelasticity of the material and adopt

$$\boldsymbol{T} = \frac{\partial \hat{\psi}}{\partial \boldsymbol{\varepsilon}^e}. \tag{21}$$

The dissipation now takes the form,

---

[ii] Whether the procedures are equivalent or not requires further analysis – our procedure does not explicitly require a minimum norm solution.

$$D = \int_R \frac{\partial \hat{\psi}}{\partial \varepsilon^e} : \dot{\varepsilon}^p \, dv = \int_R T : \dot{\varepsilon}^p \, dv. \quad (22)$$

For the purpose of the present discussion, we assume that the stress field $T$ is smooth. Then

$$D = \int_R T : \dot{\varepsilon}^p \, dv = \int_R (T_{/\!/} + T_\perp) : \dot{U}^p \, dv = \int_R T_{/\!/} : \dot{U}^p_{/\!/} \, dv + \int_R \operatorname{curl} W_T : \dot{U}^p \, dv, \quad (23)$$

where (9) has been used. Now, using (8), (12), and (13),

$$\begin{aligned}
\int_V \operatorname{curl} W_T : \dot{U}^p \, dv &= \int_V W_T : \operatorname{curl} \dot{U}^p \, dv = \int_V W_T : (-\dot{\alpha}) \, dv \\
&= \int_V W_T : \left( \operatorname{curl} \dot{U}^p - s \right) dv = \int_V T_\perp : \dot{U}^p_\perp \, dv - \int_V W_T : s \, dv,
\end{aligned} \quad (24)$$

so that (23) and (24) together imply

$$D = \int_R T : \dot{U}^p \, dv - \int_R W_T : s \, dv, \quad (25)$$

where $s = \sum_\kappa s^{(\kappa)}$. Since $\upsilon = \mathfrak{F} f$,

$$s = \sum_\kappa s^{(\kappa)} = \sum_\kappa \sum_{i=1}^{3} \sum_{j=2}^{3} S_{ij}^{(\kappa)} i_i^{(\kappa)} \otimes i_j^{(\kappa)}, \quad (26)$$

and using (13), $D$ takes the form

$$D = \int_R T : \left( \sum_\kappa \alpha^{(\kappa)} \times V^{(\kappa)} \right) dv - \int_R W_T : \left( \sum_\kappa \sum_{i=1}^{3} \sum_{j=2}^{3} S_{ij}^{(\kappa)} i_i^{(\kappa)} \otimes i_j^{(\kappa)} \right) dv. \quad (27)$$

We now recall [I] that a necessary condition for the existence of solutions to (8) is that $\operatorname{div} \alpha = 0$ (where we assume for the present discussion that $\alpha$ is sufficiently smooth). This requirement may be fulfilled by seeking the total sum of the crystallographic dislocation nucleation rates to be of the form

$$\sum_\kappa \sum_{i=1}^{3} \sum_{j=2}^{3} S_{ij}^{(\kappa)} i_i^{(\kappa)} \otimes i_j^{(\kappa)} = \operatorname{curl} \Omega \quad (28)$$

for a second order tensor $\Omega$, as can be seen from (12) and (26). Once an appropriate $\Omega$ is specified, then we use exactly the same procedure as the one used to determine $\left( f_{ij}^{(\kappa)} \right)$ from the tensor $\upsilon$ to determine $\left( S_{ij}^{(\kappa)} \right)$ from $\operatorname{curl} \Omega$. With this understanding (27) implies

$$D = \int_R \sum_\kappa X \left( T \alpha^{(\kappa)} \right) \cdot V^{(\kappa)} \, dv + \int_R -T_\perp : \Omega \, dv. \quad (29)$$

We refer to the tensor $\Omega$ as the *nucleation rate potential*.

Based on the form of (29), we now define the *driving forces* for the theory:

The driving force for the $\kappa^{th}$ slip system velocity vector ($V^{(\kappa)}$) is

$$\boldsymbol{\xi}^{(\kappa)} := \boldsymbol{X}\left(\boldsymbol{T}\boldsymbol{\alpha}^{(\kappa)}\right)^{\text{iii}}. \tag{30}$$

The driving force for the nucleation rate potential ($\Omega$) is

$$\Theta := -T_{\perp}. \tag{31}$$

We also note for later use that for arbitrary vector and second-order tensor valued functions $\boldsymbol{v}^{(\kappa)}$ and $\boldsymbol{\omega}$, the definitions

$$\left. \begin{array}{l} \boldsymbol{V}^{(\kappa)} := \text{sgn}\left(\int_{R}\sum_{\kappa^{*}} \boldsymbol{X}\left(\boldsymbol{T}\boldsymbol{\alpha}^{(\kappa^{*})}\right)\cdot \boldsymbol{v}^{(\kappa^{*})}\, dv\right)\boldsymbol{v}^{(\kappa)} \\ \boldsymbol{\Omega} := \text{sgn}\left(\int_{R} -\boldsymbol{T}_{\perp} : \boldsymbol{\omega}\, dv\right)\boldsymbol{\omega} \end{array} \right\} \Rightarrow D \geq 0. \tag{32}$$

## *6. Guidelines for Constitutive Equations arising from Theory and some Consequences*

We now examine $\boldsymbol{\xi}^{(\kappa)}$ and $\Theta$ with a view towards defining the slip system dislocation velocities and crystallographic nucleation rates. Following conventional wisdom, we adopt the point of view that constitutively specified kinetic variables be functions of their driving force fields. We intend to specify $\boldsymbol{v}^{(\kappa)}$ and $\boldsymbol{\omega}$ subject only to restrictions arising from known facts about dislocations, positive dissipation being guaranteed by the final form of constitutive equations given by (32).

### *6.1 Dislocation Velocity*

Using the symmetry of the stress tensor,

$$\boldsymbol{\xi}^{(\kappa)} = \boldsymbol{X}\left(\left\{\boldsymbol{T}\boldsymbol{i}_{p}^{(\kappa)}\right\}\otimes \boldsymbol{i}_{p}^{(\kappa)}\right)\left(\alpha_{jk}^{(\kappa)}\boldsymbol{i}_{j}^{(\kappa)}\otimes \boldsymbol{i}_{k}^{(\kappa)}\right) = \left(\boldsymbol{i}_{j}^{(\kappa)}\cdot \boldsymbol{T}\boldsymbol{i}_{p}^{(\kappa)}\right)\alpha_{jk}^{(\kappa)}\boldsymbol{i}_{p}^{(\kappa)}\times \boldsymbol{i}_{k}^{(\kappa)}, \tag{33}$$

where $\alpha_{ij}^{(\kappa)} = \boldsymbol{i}_{i}^{(\kappa)}\cdot \boldsymbol{\alpha}\,\boldsymbol{i}_{j}^{(\kappa)}$. Since $\alpha_{j1}^{(\kappa)} = 0$, (33) implies

$$\begin{aligned}\boldsymbol{\xi}^{(\kappa)} &= \left(\boldsymbol{i}_{j}^{(\kappa)}\cdot \boldsymbol{T}\boldsymbol{i}_{1}^{(\kappa)}\right)\boldsymbol{i}_{1}^{(\kappa)}\times \left(\alpha_{j2}^{(\kappa)}\boldsymbol{i}_{2}^{(\kappa)} + \alpha_{j3}^{(\kappa)}\boldsymbol{i}_{3}^{(\kappa)}\right) \\ &\quad + \left(\boldsymbol{i}_{j}^{(\kappa)}\cdot \boldsymbol{T}\boldsymbol{i}_{2}^{(\kappa)}\right)\boldsymbol{i}_{2}^{(\kappa)}\times \alpha_{j3}^{(\kappa)}\boldsymbol{i}_{3}^{(\kappa)} + \left(\boldsymbol{i}_{j}^{(\kappa)}\cdot \boldsymbol{T}\boldsymbol{i}_{3}^{(\kappa)}\right)\boldsymbol{i}_{3}^{(\kappa)}\times \alpha_{j2}^{(\kappa)}\boldsymbol{i}_{2}^{(\kappa)}. \end{aligned} \tag{34}$$

We examine the physical content of (34) by rewriting it in the form

---

[iii] Mura (1970) derives essentially the same expression in the context of a continuum without crystalline structure. His derivation, however, is not generalizable to the case when dislocation sources are included. Also, the plastic strain rate cannot be derived uniquely in his theory in terms of the dislocation density and velocity, and the same non-uniqueness translates to the driving force.

$$\begin{aligned}
\boldsymbol{\xi}^{(\kappa)} = &\left\{T_{12}^{(\kappa)}\alpha_{13}^{(\kappa)} - T_{13}^{(\kappa)}\alpha_{12}^{(\kappa)}\right\}\boldsymbol{i}_1^{(\kappa)} - T_{11}^{(\kappa)}\left\{\alpha_{13}^{(\kappa)}\boldsymbol{i}_2^{(\kappa)} - \alpha_{12}^{(\kappa)}\boldsymbol{i}_3^{(\kappa)}\right\} \\
& + \left\{T_{22}^{(\kappa)}\alpha_{23}^{(\kappa)} - T_{23}^{(\kappa)}\alpha_{22}^{(\kappa)}\right\}\boldsymbol{i}_1^{(\kappa)} - T_{21}^{(\kappa)}\left\{\alpha_{23}^{(\kappa)}\boldsymbol{i}_2^{(\kappa)} - \alpha_{22}^{(\kappa)}\boldsymbol{i}_3^{(\kappa)}\right\} \\
& + \left\{T_{32}^{(\kappa)}\alpha_{33}^{(\kappa)} - T_{33}^{(\kappa)}\alpha_{32}^{(\kappa)}\right\}\boldsymbol{i}_1^{(\kappa)} - T_{31}^{(\kappa)}\left\{\alpha_{33}^{(\kappa)}\boldsymbol{i}_2^{(\kappa)} - \alpha_{32}^{(\kappa)}\boldsymbol{i}_3^{(\kappa)}\right\},
\end{aligned} \qquad (35)$$

where $T_{ij}^{(\kappa)} = \boldsymbol{i}_i^{(\kappa)} \cdot \boldsymbol{T}\boldsymbol{i}_j^{(\kappa)}$. On considering the (common) situation when the dislocation state at a given point can be expressed as $\boldsymbol{\alpha}^{(\kappa)} = \boldsymbol{i}_2^{(\kappa)} \otimes \left(\alpha_{22}^{(\kappa)}\boldsymbol{i}_2^{(\kappa)} + \alpha_{23}^{(\kappa)}\boldsymbol{i}_3^{(\kappa)}\right)$, i.e. an infinitesimal dislocation with Burgers vector in the slip direction and line direction in the slip plane, only the second line of (35) is non-vanishing and it indicates that the in-plane component of the driving force depends on the state of stress *only* through the resolved shear stress (Schmid stress) on the relevant slip plane. The direction of the in-plane driving force is also seen to be perpendicular to the line direction $\boldsymbol{i}_1^{(\kappa)} \times \left(\alpha_{22}^{(\kappa)}\boldsymbol{i}_2^{(\kappa)} + \alpha_{23}^{(\kappa)}\boldsymbol{i}_3^{(\kappa)}\right)$, which is exactly in accord with the direction of the Peach-Koehler force on a single dislocation of classical dislocation theory. The stress-dependence of the driving force is on the total local stress which includes contributions from the stress field of the dislocation distribution as well as applied loads[iv] - this indicates that a dislocation velocity description based on a dependence on $\boldsymbol{\xi}^{(\kappa)}$ would have dislocation interactions, in the context of a linear elastic material, incorporated naturally.

Continuing with the same dislocation state, i.e. $\boldsymbol{\alpha}^{(\kappa)} = \boldsymbol{i}_2^{(\kappa)} \otimes \left(\alpha_{22}^{(\kappa)}\boldsymbol{i}_2^{(\kappa)} + \alpha_{23}^{(\kappa)}\boldsymbol{i}_3^{(\kappa)}\right)$, we note that the out-of-plane component of the driving force contains stress components other than the Schmid stress and the dependence of the dislocation velocity on such components may be construed as giving rise to non-Schmid effects. Indeed, the out-of-plane component is believed to be a necessary condition for dislocation climb and cross-slip, the latter requiring additionally a gradient in stress or dislocation density. This can be seen most simply by considering the dislocation state to consist of only a screw dislocation with Burgers vector in the slip direction. Assuming for simplicity that the velocity is proportional to the driving force, a part of the instantaneous dislocation density evolution is $-curl\left(\left\{\alpha_{22}^{(\kappa)}\boldsymbol{i}_2^{(\kappa)} \otimes \boldsymbol{i}_2^{(\kappa)}\right\} \times -T_{23}^{(\kappa)}\alpha_{22}^{(\kappa)}\boldsymbol{i}_1^{(\kappa)}\right)$ (up to the proportionality constant) which may be expressed as

---

[iv] This is to be contrasted with a dependence on the so-called 'defect stress' (Gurtin, 2001).

$$\left[-curl\left(\left\{\alpha_{22}^{(\kappa)}\boldsymbol{i}_{2}^{(\kappa)}\otimes\boldsymbol{i}_{2}^{(\kappa)}\right\}\times-T_{23}^{(\kappa)}\alpha_{22}^{(\kappa)}\boldsymbol{i}_{1}^{(\kappa)}\right)\right]_{ri}=e_{ijk}\left\{\alpha_{22}^{(\kappa)2}T_{23}\right\}_{,j}\delta_{r2}\delta_{k3}$$

$$\Rightarrow\left[-curl\left(\left\{\alpha_{22}^{(\kappa)}\boldsymbol{i}_{2}^{(\kappa)}\otimes\boldsymbol{i}_{2}^{(\kappa)}\right\}\times-T_{23}^{(\kappa)}\alpha_{22}^{(\kappa)}\boldsymbol{i}_{1}^{(\kappa)}\right)\right]_{21}=\left\{\alpha_{22}^{(\kappa)2}T_{23}\right\}_{,2}. \tag{36}$$

The above suggests a reassuring similarity with the physical picture of cross slip occurring from a gliding screw dislocation that meets a pair of obstacles along its line direction (reflected in the stress gradient required along the line direction in (36), assuming a straight, cylindrical dislocation with no gradient in density along the line direction) and then forms a pair of edge jogs on the cross slip plane (the development of 21 component in the dislocation density rate in (36) ). It is indeed tempting here to postulate that a necessary condition for cross slip is the alignment of the driving force with a slip plane of the material and the presence of gradients in the $T_{23}$ component along the slip direction in the slip system basis.

Other non-Schmid behaviors can also be seen to arise whenever the dislocation state at a point may be represented as $\boldsymbol{\alpha}^{(\kappa)}=\left(\sum_{i=1}^{3}k_{i}\boldsymbol{i}_{i}^{(\kappa)}\right)\otimes\left(\sum_{i=2}^{3}l_{i}\boldsymbol{i}_{i}^{(\kappa)}\right)$, where $k_{i},l_{i}$ are scalars. This state represents a dislocation with line direction in the slip plane and Burgers vector in a direction not necessarily in the slip plane. Such a dislocation state may arise in the modeling of immobile or relatively immobile dislocations, e.g Lomer-Cottrell lock, screw dislocations in bcc materials at low temperatures with core structure[v]. For a screw dislocation with its core spread out so that it has some edge component, the driving force can be seen to contain the stress component $T_{31}^{(\kappa)}$ which is believed to give rise to the orientation dependence of yield in some intermetallic compounds. The driving force in any of these cases contains non-Schmid stress components.

It is also worthy of note here that in all instances the driving force incorporates a dependence on the dislocation state naturally. Since the dislocation density is a solution variable of the theory, such a dependence does not have to be further phenomenologically modeled, the latter being the typical case in slip based conventional crystal plasticity where dislocation density is not a solution variable.

All of the above features appear to be desirable attributes of a theory of continuum dislocation mechanics. In analogy with conventional ideas related to the motion of discrete dislocations, we would now want to adopt the rule that the in-plane and out-of-plane components of the slip system dislocation velocity vector be functions of the corresponding components of the driving force, i.e.

---

[v] Of course, dealing with core structure also requires, in addition to the above kinematics, a consideration of nonlinear crystal elasticity incorporating lattice symmetries and periodicity.

$$\begin{aligned} \boldsymbol{V}_n^{(\kappa)} &:= \left(\boldsymbol{V}^{(\kappa)} \cdot \boldsymbol{i}_1^{(\kappa)}\right) \boldsymbol{i}_1^{(\kappa)} = function\left(\boldsymbol{\xi}^{(\kappa)} \cdot \boldsymbol{i}_1^{(\kappa)}\right) \boldsymbol{i}_1^{(\kappa)} \\ \boldsymbol{V}^{(\kappa)} - \boldsymbol{V}_n^{(\kappa)} &= function\left(\boldsymbol{\xi}^{(\kappa)} - \left\{\boldsymbol{\xi}^{(\kappa)} \cdot \boldsymbol{i}_1^{(\kappa)}\right\} \boldsymbol{i}_1^{(\kappa)}\right). \end{aligned} \quad (37)$$

However, before taking such a step we have to ensure that the theory represents two fundamental facts about dislocations: a straight discrete dislocation does not move under its own stress field in an infinite homogeneous medium; and a straight dislocation in such a medium should move with constant velocity under a spatially uniform applied stress field - of course, only in the geometrically linear theory with a linear elastic constitutive law for stress does it make sense to speak of an 'applied' stress field.

In the present continuum theory, a straight dislocation may be modeled as a cylinder of a certain core radius $r_0$ (to be specified external to the theory) with a non-vanishing dislocation density in it [I]. The integral of the dislocation density field over the cross-section of the cylinder represents the strength of the dislocation. Let the dislocation velocity be chosen according to (37), with the functional relationship being a simple proportionality between velocity and driving force. Then, using the solution for a screw dislocation of strength $b$ derived in [I] for an otherwise arbitrary axisymmetric density distribution in the core cylinder, a simple calculation reveals that under the above constitutive assumption for dislocation velocity the dislocation cylinder 'disintegrates' under its own stress field, the driving force tending to spread it out in the domain[vi]. Such a result clearly is physically unacceptable, and this observation leads us to modify our definition of the dislocation velocity.

For fixed $\kappa, i, j$, let $\boldsymbol{P}_{ij}^{(\kappa)}$ be the sum of terms in $\boldsymbol{\xi}^{(\kappa)}$ (35) that are linear in $\alpha_{ij}^{(\kappa)}$ with direction orthogonal to $\boldsymbol{i}_1^{(\kappa)}$. Let $\boldsymbol{N}_{ij}^{(\kappa)}$ be the sum of terms linear in $\alpha_{ij}^{(\kappa)}$ with direction parallel to $\boldsymbol{i}_1^{(\kappa)}$. We denote by $\hat{\boldsymbol{P}}_{ij}^{(\kappa)}(\boldsymbol{x};\boldsymbol{y})$ and $\hat{\boldsymbol{N}}_{ij}^{(\kappa)}(\boldsymbol{x};\boldsymbol{y})$ the values of the $\boldsymbol{P}_{ij}^{(\kappa)}$ and $\boldsymbol{N}_{ij}^{(\kappa)}$ fields at the point $\boldsymbol{x}$ of an infinite, homogeneous medium of crystal symmetry given by that of the actual crystal at the point $\boldsymbol{y}$, and generated from the presence of a straight cylindrical dislocation of core radius $r_0$ with axis passing

---

[vi] As known to Nabarro (1987) p 42.

through $y$ whose line points in the direction corresponding to the indices $(j,\kappa)$ and whose Burgers vector is in the direction corresponding to the indices $(i,\kappa)$. These fields can be generated as solutions to (8), (10), and (11) for the infinite domain and dislocation density distribution defined above. Let $A_j^{(\kappa)}(y)$ be the set of points contained in the closed disc of radius $2r_0$ with center $y$, perpendicular to the direction corresponding to the indices $(j,\kappa)$. We now define

$$\bar{\hat{P}}_{ij}^{(\kappa)}(y) = \int_{A_j^{(\kappa)}(y)} \hat{P}_{ij}^{(\kappa)} \, da \quad ; \quad \bar{\hat{N}}_{ij}^{(\kappa)}(y) = \int_{A_j^{(\kappa)}(y)} \hat{N}_{ij}^{(\kappa)} \, da, \qquad (38)$$

$$\bar{\xi}^{(\kappa)}(y) = \sum_{i=1}^{3}\sum_{j=2}^{3}\left\{\left(\int_{A_j^{(\kappa)}(y)} P_{ij}^{(\kappa)} \, da - \bar{\hat{P}}_{ij}^{(\kappa)}(y)\right) + \left(\int_{A_j^{(\kappa)}(y)} N_{ij}^{(\kappa)} \, da - \bar{\hat{N}}_{ij}^{(\kappa)}(y)\right)\right\}, \qquad (39)$$

$$v_n^{(\kappa)} := f_n\left(\bar{\xi}^{(\kappa)} \cdot i_1^{(\kappa)}\right) i_1^{(\kappa)} \quad ; \quad f_n(0) = 0$$

$$v_p^{(\kappa)} := f_p\left(\left|\bar{\xi}^{(\kappa)} - \left\{\bar{\xi}^{(\kappa)} \cdot i_1^{(\kappa)}\right\} i_1^{(\kappa)}\right|\right) \frac{\bar{\xi}^{(\kappa)} - \left\{\bar{\xi}^{(\kappa)} \cdot i_1^{(\kappa)}\right\} i_1^{(\kappa)}}{\left|\bar{\xi}^{(\kappa)} - \left\{\bar{\xi}^{(\kappa)} \cdot i_1^{(\kappa)}\right\} i_1^{(\kappa)}\right|} \quad ; \quad f_p(0) = 0 \qquad (40)$$

$$V^{(\kappa)} = v_n^{(\kappa)} + v_p^{(\kappa)},$$

with $V^{(\kappa)}$ to be defined according to (32). We note here that while the physical dimensions of $\xi^{(\kappa)}$ is a force per unit volume, $\bar{\xi}^{(\kappa)}$ has units of force per unit length. Essentially, the above choice for the constitutive equation has been designed to ensure that a single, straight dislocation in an infinite medium remains stationary under its own stress field[vii] and, under the action of a homogeneous applied stress field, moves as a rigid body. That the latter is a consequence of the theory can be deduced by considering the solution for the motion of a single straight dislocation under spatially uniform velocity field derived in [I] and noting that (40) dictates that the driving force at every point of the cylinder representing a straight dislocation under a homogeneous applied stress field is identical.

The functions $f_p$ and $f_n$ will, in general, also depend upon a dislocation drag coefficient, temperature, and the slip system dislocation density tensor. The latter may be necessary to phenomenologically model additional resistance to dislocation motion if the dislocation state contains Burgers vectors off of the slip plane. As a

---

[vii] While the above procedure is essential to adopt in a theory where the elasticity is linear, it is reasonable to expect that a proper accounting for nonlinear crystal elasticity would make the above construction unnecessary (Nabarro, 1987).

practical device, the $\overline{\hat{P}}_{ij}^{(\kappa)}$ and $\overline{\hat{N}}_{ij}^{(\kappa)}$ terms may also be modeled by phenomenological resistances.

As for the question of work-hardening due to short range interactions, we note that individual dislocation stress fields in this theory are not singular for smooth dislocation density distributions within the core, as shown in [I], while reproducing the classical elastic fields outside the core region. Consequently, even for dislocations at very close proximity, the theory would be capable of representing the effects of elastic interactions of dislocations, at least in principle, up to the sophistication in the elastic constitutive law. Of course, all of the ideas presented have their analogues in a theory employing nonlinear crystal elasticity and finite deformations, and in such a theory a more precise accounting of such interactions would be possible. In practice, say in the context of numerical computation, it would perhaps be unreasonable to demand mesh resolution on the order of the interatomic spacing when dealing with a physical situation representative of a large collection of dislocations[viii]. In such a case, the constitutive equation for dislocation velocity may be modified for the effects of computationally unresolved short-range interactions by modeling insights as presented in the treatise of Kocks, Argon, and Ashby (1975). It may also be hoped that a 'complete', i.e. including field equations and constitutive equations, coarse scale theory may be developed by the Method of Invariant Manifolds (Muncaster, 1983 a,b) by averaging the fine-scale theory proposed in this work and [I], in which case the effects of short range interactions in the macroscopic theory should be accounted for naturally. This is an exciting and promising research direction since the Method of Invariant Manifolds was developed as a generalization of the procedure used to derive, in a precise and special sense, the field equations and constitutive equations of fluid dynamics from those of the kinetic theory.

*6.2 Nucleation Rate*

As opposed to the dislocation velocity, much less detail seems to be apparent from theory for the constitutive equation for nucleation rate $s$. The simplest possibility is to assume

$$\boldsymbol{\Omega} = -c\boldsymbol{T}_\perp, \tag{41}$$

---

[viii] This may be reasonable in the case of dealing with a collection of a few dislocations, as may occur in studies of semiconductor thin films grown epitaxially.

where $c$ is a scalar parameter that is required on dimensional grounds. If $c$ is assumed to be a material constant, the choice (41) implies

$$\mathbf{s} = -c \, curl \, \mathbf{T} \,, \tag{42}$$

which is true due to the fact $curl \, \mathbf{T}_{//} = \mathbf{0}$.

While (42) is a sound idea from the thermodynamic point of view, it does not appear to have a simple mechanically intuitive interpretation related to dislocation nucleation. One mechanical implication is the following (essentially) force sum

$$\int_A \mathbf{s} \, \mathbf{n} \, da = -c \int_A curl \, \mathbf{T} \, \mathbf{n} \, da = -c \oint_{C^*} \mathbf{T} \, d\mathbf{x} \,, \tag{43}$$

for any surface $A$ bounded by the closed curve $C^*$, but how this may be precisely related to dislocation nucleation is not clear. This in itself is not necessarily a shortcoming of a constitutive equation since thermodynamic driving forces do not always have interpretations that appeal to mechanically guided intuition – that a screw dislocation, viewed as mechanical entity, should move perpendicular to itself under the action of an applied shear stress in the direction of its line defies all mechanical notions of applied 'mechanical' force and resulting motion. Be that as it may, much experimental work is required before accepting (42) as a legitimate constitutive equation for nucleation. For instance, it would imply that dislocation nucleation could occur in the presence of an inhomogeneous hydrostatic stress field which is not curl free – whether dislocation nucleation can happen at all under hydrostatic stress fields is a fact that can perhaps be used to check the validity of (42). We indulge in this speculative vein only to point out that the curl of the stress field may have a connection to dislocation nucleation, especially since it emerges from the same thermodynamic procedure that delivers a driving force for dislocation velocity that is in accord with a number of physically observed facts about dislocations.

Before ending this section, we note that (42) may be interpreted as indicating that the stress for dislocation nucleation is nonlocal in the dislocation nucleation rate. A rate-independent counterpart of such a conclusion (which is achievable with a suitable rate-independent choice in place of the parameter $c$) seems to be in superficial accord with the broad conclusion of an atomistic-dislocation theory study (Miller *et al.*, 1998) that indicates that a nonlocal-in–dislocation-density stress criterion for nucleation is a better model for nucleation than one based on a local interplanar potential, but a more precise study in this regard is warranted.

*6.3 Some Consequences*

We now discuss the notions of steady dislocation microstructures and yielding within the context of this theory, without further commitment to particular constitutive assumptions beyond (40) and (42). Steady (constant in time) dislocation density distributions under no applied loads are important predictions since they can be put to test against experimental observations. The importance of the concept of yielding needs no elaboration for an audience well-versed in plasticity.

A steady microstructure is a state of the body where $\dot{\boldsymbol{\alpha}}^{(\kappa)} \equiv \boldsymbol{0}, \forall \kappa$. The field equations of the theory indicate that it is possible to obtain stressed, steady microstructures with time-varying total deformation under no applied loads. This can happen as follows: let there be no applied loads but a non-trivial stress field due to the presence of dislocations. Additionally, let this stress field and the dislocation distribution be such that we have a steady microstructure instantaneously. Even though a non-trivial stress field results in the evolution of the $\tilde{\boldsymbol{U}}^p$ field in general, it is possible for the instantaneous steady microstructure to persist in time since it can be seen that the solution for the stress field in (8)-(11) under no applied loads does not vary in time with the $\tilde{\boldsymbol{U}}^p$ field, the only evolution being that of the total deformation which 'makes up' for the compatible part of the evolving $\tilde{\boldsymbol{U}}^p$ field. Additionally, we observe that the stress field in the case of a steady microstructure under no applied loads is a functional only of the geometry and elasticity of the body along with the total dislocation density field on it, and consequently with the assumed constitutive structure given by (40) and (42), the conditions for a steady microstructure can be entirely expressed as a set of nonlinear, (spatial) integro-partial differential equations in the slip system dislocation density fields. The set of all possible solutions to this set of equations on a given body characterizes the entire class of steady microstructures under no applied loads for that particular body. Such a characterization may be undertaken by the methods of Lie Group theory.

It is natural to ask at this point as to what progress can be made on the above question without attempting an exhaustive classification. A large class of solutions are contained in dislocation density distributions that result in no stress in the body – in addition, these are also equilibrium solutions of theory. It is easy to see from the field equations that, in the absence of applied loads, whenever the total dislocation density field on the body can be represented as a curl of a skew symmetric tensor field we have a zero-stress dislocation density distribution. In particular a homogeneous total dislocation density distribution satisfies this condition, as can be seen by solving (8) corresponding to this density distribution by the method of solution of exterior differential equations illustrated in [I] and observing that the distortion field so obtained has a compatible strain field. Alternatively, Kröner's solution method for the elastic theory of dislocations (Kröner, 1981) indicates directly that non-trivial stress fields are obtained only if the dislocation density field is inhomogeneous. As an aside, it would be interesting to understand how an *equilibrium* homogeneous dislocation density field can be rationalized as being a limit of some distribution of discrete, cylindrical dislocation curves in the body.

It is important to note here that observed dislocation cell structures, i.e. three dimensional regions free of dislocation density separated by dislocation walls, can be modeled as a rigorous stress-free microstructure in the present theory as regions of uniform orientation represented by locally uniform skew symmetric elastic distortion field separated by layers over which the skew symmetric elastic distortion field transitions from one orientation to another. This observation was first made by Head *et al.* (1993) in their study of an equilibrium theory of dislocation mechanics. Of course, it is of equal, if not more, interest to know how such a microstructure *develops* under any particular loading program in a general nonequilibrium process. The present theory is equipped to deal with this question in dislocation mechanics too – albeit, perhaps, only with the aid of approximate numerical computations.

Finally, the idea that a reasonable model of macroscopic yielding may well be achievable within the theory can be inferred from the arguments presented in Head *et al.* (1993) that also applies to the present discussion. Essentially, if one considers a stress free cellular dislocation microstructure as described above as an initial state and considers the application of loads to the body, for small loads the cellular structure persists and dislocation motion (slip deformation evolution in the theory) takes place only in the walls. Since the walls form a very small volume fraction of the total material, macroscopically this wall plasticity is not discernible until a point where the applied load is high enough to make the wall structure disintegrate and the dislocation density evolution takes place even in the cell interiors. The critical load at which this transition takes place may be interpreted as the yield point of the material.

Apart from providing guidance on the nature of some predictions, the rigorous conclusions and plausible conjectures presented in this subsection form precise targets for computational approaches based on the theory.

## *7. Boundary Conditions on Dislocation Density and Closure Condition*

In [I], it is shown through a simple example that boundary conditions are required for (12) on finite domains, if we demand that an adequately posed general theory should provide for unique solutions to the equation for dislocation density evolution in the uncoupled case, i.e. the dislocation velocity is not a function of stress and, in particular, when it is a constant. Relying on heuristic arguments to obtain some idea of conditions required to have a closed theory, we observe that (8)-(9) has at most one solution, by design, if the dislocation density field is specified. From experience with the mathematical structure of phenomenological plasticity we know that (10)-(11) has at most one solution under the usual traction/displacement/mixed boundary conditions for the equilibrium equations when $U^p$ is known. Noticing that (13) is essentially an ordinary differential equation, we proceed on the assumption that specifying initial conditions for it is sufficient for a well-defined evolution. This leaves conditions of closure for (12) to be derived. That this is the only condition that is required may be further substantiated in the case when the dislocation velocities and nucleation rates are considered to be specified functions of space and time.

One option in deriving such a condition is to consider the specification of initial and boundary conditions. It is natural to think of specifying initial conditions on the dislocation density fields; however, the precise nature of any boundary conditions that may be required is not obvious. Closure can also be ensured in the case of some partial differential equations without specifying boundary conditions,e.g. (8)-(9). However, it has to be made sure that the conditions prescribed are not overly restrictive so as to preclude physical behaviors.

In the context of dislocation density evolution, one such behavior we have to be concerned about is the increase in dislocation line length in the body. Frank-Read sources and expanding dislocation loops are believed to be some of the main mechanisms behind the increase in total dislocation density by several orders of magnitude in a cold worked material. It is important that the general theory be capable of predicting such behavior, and any uniqueness condition that is specified not preclude such growth in the dislocation density.

With the above ideas in mind, we begin with two simple examples that illustrate the capability of the theory to model growth in dislocation line length, and consequently in the magnitude of the dislocation density, in the body. We then attack the problem of uniqueness of solutions through the specification of initial conditions and appropriate boundary conditions.

*7.1 Initiation of Bowing of a Screw Segment*

In [I], the equations governing the evolution of a dislocation density field of the form $b \otimes t$, where $b$ is a spatially uniform vector field (Burgers vector) and $t$ is a vector field that lies on a slip system (the line direction field), have been derived under the simplifying assumption that there exists only one slip system. These equations are

$$\dot{t}_1 = 0$$
$$\dot{t}_2 = \{V|t|\}_{,3} \qquad (44)$$
$$\dot{t}_3 = \{-V|t|\}_{,2},$$

where components and coordinates are with respect to the slip system basis, $V$ is the dislocation velocity magnitude, and the dislocation velocity is assumed to be in the slip plane and normal to the direction $t$.

Suppose the fields $V$ and $t$ vary at most with $x_2$ at the initial instant, and let the initial condition represent a straight, cylindrical screw dislocation in the slip direction with no variation of dislocation density in the cylinder along $x_2$. Under these circumstances, only the last of (44) is non-trivial and takes the form

$$\dot{t}_3 = -V_{,2} t_2. \qquad (45)$$

If we now think of a pinned segment, i.e. a dislocation velocity variation along the axis of the dislocation that vanishes outside a certain segment and is a symmetric parabola within the segment, then the center of the segment remains screw in character while the maximum increments in edge character appear at the pinning points with opposite signs, just as physically expected in the bowing of a screw segment in a Frank-Read source. There is a great deal of similarity in the kinematics of the above situation and that of cross slip discussed in Section 6.1, up to the orientation of the plane in which bowing takes place. One important difference is that if the velocity was assumed to be proportional to the driving force, then bowing in the slip plane occurs due to a gradient in the resolved shear stress ($T_{21}$ component) in the $x_2$ direction, whereas cross slip requires a gradient in the $T_{23}$ component in the $x_2$ direction.

## 7.2 Expansion of a Polygonal Loop

We would now like to derive a two-dimensional solution to (44) as opposed to the solution for the motion of a straight dislocation under constant velocity [I] which was essentially one-dimensional. It turns out that one of the simplest solutions that can be derived corresponds to the expansion of a polygonal dislocation loop.

We assume that $t_1 \equiv 0$ and $V$ a constant. We define the variable

$$s = Vt. \tag{46}$$

Even though the field $t$ can depend upon $x_1$, and necessarily does when modeling a dislocation segment on the slip plane by a cylinder, such a dependence is only parametric and we consider the problem in $x_2 - x_3 - s$ space. With respect to these variables, (44) takes the form

$$\begin{aligned} t_{2,s} &= \varphi_{,3} \\ t_{3,s} &= -\varphi_{,2}, \end{aligned} \tag{47}$$

where $\varphi \equiv \sqrt{t_2^2 + t_3^2}$. If we now consider a vector field $T = (-t_3, t_2, \varphi)$ in this space and demand that

$$\operatorname{curl} T := (\varphi_{,3} - t_{2,s}, -\varphi_{,2} - t_{3,s}, t_{2,2} + t_{3,3}) = (0, 0, 0), \tag{48}$$

then we have a solution to (47). But if $\operatorname{curl} T$ vanishes, then it can be represented as a gradient of a scalar field:

$$\begin{aligned} \theta_{,2} &= -t_3 \\ \theta_{,3} &= t_2 \\ \theta_{,s} &= \varphi. \end{aligned} \tag{49}$$

Clearly, if we can find a scalar function $\theta$ satisfying (49), then (48) is satisfied and hence (47).

Assuming there exists a solution $\theta$ of (49), we note that

$$(\theta_{,s})^2 = (\theta_{,2})^2 + (\theta_{,3})^2. \tag{50}$$

If we further assume that $\theta$ is of the form

$$\theta(x_2, x_3, s) = f(m_2 x_2 + m_3 x_3 - cs - a), \tag{51}$$

where $f$ is a function of a real variable, and $m_2, m_3, c, a$ are constants, then (50) implies

$$c = \pm\sqrt{m_2^2 + m_3^2}. \tag{52}$$

We are now in a position to explore the evolution of a polygonal dislocation loop. For initial conditions, we assume the plane of the loop to be parallel to the $x_2 - x_3$ plane. The loop is visualized as straight cylindrical segments joined in the shape of a regular polygon. The polygon formed by joining the axes of these segments is assumed to lie in the plane $x_1 = 0$, with center at the origin. We assume that the polygon is $n$-sided, and arbitrarily number its $n$ triangular sectors consecutively from 1 to $n$. Let the direction cosines, with respect to the $x_2$ and $x_3$ directions, of the in-plane normal to the axis of the dislocation in the $i^{th}$ sector be $(m_2^i, m_3^i)$. The sense of the normal is meant in the outward direction w.r.t. the origin. The initial condition is now defined as

$$\left. \begin{array}{l} t_2(x_1, x_2, x_3, 0) = -m_3^i \, \beta(x_1, m_2^i x_2 + m_3^i x_3 - a) \\ t_3(x_1, x_2, x_3, 0) = m_2^i \, \beta(x_1, m_2^i x_2 + m_3^i x_3 - a) \end{array} \right\} \text{ for } (x_2, x_3) \text{ in the } i^{th} \text{ sector.} \quad (53)$$

The function $\beta(x_1, \bullet)$ represents the variation of the dislocation density within the core cylinder for fixed $x_1$. It is assumed to be smooth in $(-\infty, \infty)$, and zero everywhere in $(-\infty, \infty)$ except in the interval $[-w(x_1), w(x_1)]$.

We now define

$$\left. \begin{array}{l} t_2(x_1, x_2, x_3, t) = -m_3^i \, \beta(x_1, m_2^i x_2 + m_3^i x_3 - Vt - a) \\ t_3(x_1, x_2, x_3, t) = m_2^i \, \beta(x_1, m_2^i x_2 + m_3^i x_3 - Vt - a) \end{array} \right\} \text{ for } (x_2, x_3) \text{ in the } i^{th} \text{ sector,} \quad (54)$$

and note that (54) is a solution to (44) in the interior of the sectors, assuming $t_1 \equiv 0$ for all times. It is easily seen that the solution corresponds to an expanding loop with the straight dislocation segment in each sector translating outwards, w.r.t. the origin, with constant velocity $V$.

A slight complication arises if we interpret the governing equations in the strong form - for fixed $x_1$, the solution is not differentiable on the planes that divide the $n$ sectors in $x_2 - x_3 - s$ space. We get around this difficulty by posing the problem

$$curl\, T = (0, 0, 0) \quad (55)$$

in variational form with continuous test functions that are piecewise smooth. A consequence of this is that if $T$ is smooth and satisfies the strong form (55) in regions other than a finite number of internal surfaces, and on these surfaces its tangential component is continuous, then the weak form of the problem is satisfied.

It is clear that the required condition in the interior of the sectors is satisfied. As for the continuity condition on the internal surfaces dividing the sectors, we note that if we define the function

$$\lambda(x_1, x_2, x_3, s) = -\int_0^{(m_2^i x_2 + m_3^i x_3 - s - a)} \beta(x_1, p) dp \text{ for } (x_2, x_3) \text{ in the } i^{th} \text{ sector} \quad (56)$$

then, for fixed $x_1$,

$$T = grad\, \lambda \text{ (in } x_2 - x_3 - s \text{ space) for } (x_2, x_3) \text{ in the interior of each sector.} \quad (57)$$

Since $\lambda$ is a continuous function, for fixed $x_1$ its tangential derivative is necessarily continuous on any surface in $x_2 - x_3 - s$ space (by a theorem of Maxwell), and we are done with proving that (54) is a legitimate weak solution to (44) with initial conditions defined by (53) and $t_1 \equiv 0$.

### 7.3 Uniqueness of dislocation density evolution allowing growth

We now concern ourselves with deducing a sufficient condition for assuring uniqueness of solutions to a system of the form

$$\dot{\boldsymbol{\mu}} = -curl(\boldsymbol{\mu} \times \boldsymbol{V}) + \boldsymbol{r}; \; \boldsymbol{\mu}(\boldsymbol{x}, 0) = \boldsymbol{\mu}_0(\boldsymbol{x}) \text{ on } R;. \quad (58)$$

where $\boldsymbol{\mu}$ and $\boldsymbol{r}$ are second-order tensor fields representative of the slip system dislocation density tensors and slip system source, and $\boldsymbol{V}$ is a vector field representing the slip system dislocation velocity. $\boldsymbol{V}$ and $\boldsymbol{r}$ are assumed to be prescribed functions of space and time so that the system is linear (with variable coefficients). The latter assumption is far from physical reality as we have already seen, but we proceed on the belief arising from experience with other partial differential equations that the nature of boundary conditions do not change in making the transition from the variable coefficient linear case to a quasilinear or fully nonlinear partial differential equation that may arise due to complexity in coefficients.

Let a $\boldsymbol{\mu}$ exist that satisfies (58). Then

$$\begin{aligned}
\int_R \mu_{ij} \dot{\mu}_{ij} \, dv &= -\int_R \mu_{im} \left(\mu_{im} V_n\right)_{,n} dv + \int_R \mu_{in} \left(\mu_{im} V_n\right)_{,m} dv + \int_R \mu_{ij} r_{ij} \, dv \\
&= -\int_R \left(\frac{1}{2} \mu_{im} \mu_{im} V_n\right)_{,n} dv - \int_R \frac{1}{2} \mu_{im} \mu_{im} V_{n,n} \, dv + \int_R \mu_{in} \mu_{im} V_{n,m} \, dv \quad (59) \\
&\quad + \int_R \mu_{in} \mu_{im,m} V_n \, dv + \int_R \mu_{ij} r_{ij} \, dv.
\end{aligned}$$

Let $\partial R_i$ be the inflow part of the boundary on which $\boldsymbol{V} \cdot \boldsymbol{n} < 0$ and $\partial R_o$ be the outflow part on which $\boldsymbol{V} \cdot \boldsymbol{n} > 0$, where $\boldsymbol{n}$ is the outward unit normal to the boundary. Then (59) implies

$$\frac{d}{dt}\int_{R}\mu_{ij}\mu_{ij}\,dv = \int_{\partial R_i}\mu_{im}\mu_{im}|V_s n_s|\,da - \int_{\partial R_o}\mu_{im}\mu_{im}|V_s n_s|\,da$$
$$+ 2\int_{R}\mu_{in}\mu_{im}V_{n,m}\,dv - \int_{R}\mu_{im}\mu_{im}V_{n,n}\,dv \qquad (60)$$
$$+ 2\int_{R}\mu_{in}\mu_{im,m}V_n\,dv + 2\int_{R}\mu_{ij}r_{ij}\,dv.$$

Nye (1953) defined a dislocation density tensor as a measure of Burgers vector per unit area, of the normal component of the dislocation lines threading the area. If we multiply and divide this areal density by the line length along the normal to the infinitesimal element of area under consideration, then the measure can as well be interpreted as a Burgers-vector-times-line-length per unit volume measure of dislocation density. With this interpretation, (60) has a natural physical meaning. It suggests that the field equation for dislocation density evolution implies that the rate of change of the square of the magnitude of the volumetric dislocation density in the body equals the net flux of the quantity through the boundaries plus terms that reflect *growth* and annihilation depending upon the nature of the dislocation velocity, source, and the gradients of the dislocation density field on the body.

Let us consider two solutions, $\mu_1$ and $\mu_2$, of (58) and denote their difference by $\rho$. Then $\rho$ satisfies

$$\dot{\rho} = -\mathrm{curl}(\rho \times V) \quad ; \quad \rho(x,0) = \mathbf{0} \text{ on } R \qquad (61)$$

as well as

$$\mathrm{div}\,\rho = \mathbf{0} \text{ on } R. \qquad (62)$$

Using (62) along with a procedure similar to the one used to derive (60), we have

$$\frac{d}{dt}\int_{R}\rho_{ij}\rho_{ij}\,dv = \int_{\partial R_i}\rho_{im}\rho_{im}|V_s n_s|\,da - \int_{\partial R_o}\rho_{im}\rho_{im}|V_s n_s|\,da$$
$$+ 2\int_{R}\rho_{in}\rho_{im}V_{n,m}\,dv - \int_{R}\rho_{im}\rho_{im}V_{n,n}\,dv. \qquad (63)$$

At this juncture it is possible to demand a global negative-semidefiniteness criterion of the dislocation velocity field (in the domain and the inflow boundary) that ensures that the LHS of (63) is non-positive for all times, thus guaranteeing uniqueness. However, this would be inadequate for the problem being considered for two reasons: it precludes growth in the magnitude of the dislocation density, and such a criterion would be hopelessly impractical to demand of a stress and dislocation density dependent velocity field, which is the actual situation to which we would like our conclusions to apply.

Instead we demand that, in addition to (58), the following **boundary condition** should also be satisfied by all solutions:

$$\mu(\mathbf{V}\cdot\mathbf{n})=\mathbf{F} \text{ on } \partial R_i, \tag{64}$$

where $\mathbf{F}$ is a prescribed *inward* flux (second-order tensor), so that

$$\rho_{im}V_s n_s = -\rho_{im}|V_s n_s|=0 \text{ on } \partial R_i. \tag{65}$$

Using (65), we have

$$\frac{d}{dt}\int_R \rho_{ij}\rho_{ij}\,dv \leq 2\int_R \rho_{in}\rho_{im}V_{n,m}\,dv - \int_R \rho_{im}\rho_{im}V_{n,n}\,dv$$
$$\leq \int_R 2|\rho_{in}\rho_{im}V_{n,m}|\,dv + \int_R \rho_{im}\rho_{im}|V_{n,n}|\,dv. \tag{66}$$

Henceforth, we use the notation

$$\rho(t):=\int_R \rho_{ij}(\mathbf{x},t)\rho_{ij}(\mathbf{x},t)\,dv. \tag{67}$$

Assuming $\mathbf{V}$ and $grad\,\mathbf{V}$ to be continuous functions on the closed and bounded domain $R\cup\partial R$, there exists a non-negative function of time $M_1$ defined by

$$\max_{\mathbf{x}\in R\cup\partial R}|V_{n,n}(\mathbf{x},t)|=M_1(t) \tag{68}$$

so that

$$\int_R \rho_{im}\rho_{im}|V_{n,n}|\,dv \leq M_1(t)\rho(t). \tag{69}$$

Next we estimate the first term on the RHS of (66). We think of $\rho_{im}$ as a $9\times 1$ column vector $X_I$ and $\delta_{ri}V_{n,m}$ as a $9\times 9$ matrix $A_{IJ}$ so that

$$\rho_{in}\rho_{im}V_{n,m} = \rho_{rn}\delta_{ri}V_{n,m}\rho_{im} = X_I A_{IJ} X_J. \tag{70}$$

The Cauchy-Schwarz inequality in $\mathbb{R}^9$ implies that for each $I$

$$(A_{IJ}X_J)^2 \leq (A_{IJ}A_{IJ})(X_L X_L) \text{ no sum on } I. \tag{71}$$

Let

$$\max_{\mathbf{x}\in R\cup\partial R} A_{IJ}(\mathbf{x},t)A_{IJ}(\mathbf{x},t)=K_I(t), \text{ no sum on } I, \text{ and } \max_{I=1,9} K_I(t)=M_2(t). \tag{72}$$

Another application of the Cauchy-Schwarz inequality yields

$$|X_I A_{IJ} X_J|^2 \leq (A_{IJ}X_J A_{IK}X_K)(X_L X_L) \leq 9M_2(t)(X_L X_L)^2, \tag{73}$$

so that there exists a non-negative function of time $M_3$ such that

$$\int_R 2|\rho_{in}\rho_{im}V_{n,m}|\,dv \leq M_3(t)\rho(t). \tag{74}$$

Equations (66), (69), and (74) imply

$$\frac{d}{dt}\rho(t) \leq M(t)\rho(t) \quad ; \quad M(t):=M_1(t)+M_3(t). \tag{75}$$

Consequently,

$$\rho(t^*) \leq \rho(0) + \int_0^{t^*} M(t)\rho(t)dt, \tag{76}$$

and, since $M(t) \geq 0, \rho(0) \geq 0$, using the Gronwall inequality

$$0 \leq \rho(t^*) \leq \rho(0)\exp\left[\int_0^{t^*} M(t)dt\right] \Rightarrow \rho(t^*) = 0, \tag{77}$$

since $\rho(0) = 0$. Hence we have proved that $\boldsymbol{\mu}_1(\boldsymbol{x},t) = \boldsymbol{\mu}_2(\boldsymbol{x},t)$ almost everywhere on $R$ for all $t$ in any finite interval of time.

We note that the initial condition and boundary condition play a crucial role in the uniqueness proof, as expected. It is also worthy of note that the deduced boundary condition is *not* the same as would result by considering the variational form of (58) and examining the resulting boundary term, as is often done to explore natural boundary conditions corresponding to governing partial differential equations. It is to be noted here that depending upon the nature of the dislocation velocity field on the boundary, there may be instances when *no boundary condition need be specified*, e.g. if the entire boundary is an outflow surface. The boundary condition has a direct physical meaning as a surface flux of the dislocation density (in the volumetric interpretation as discussed above). It can also be seen that if different velocity vectors are associated with the evolution of the various edge and screw components (w.r.t. the slip system basis) of the slip system dislocation density tensor, then significant complications arise in the uniqueness proof and it does not go through in the present form.

When the boundary condition (64) is applied to the actual case of a dislocation density and stress dependent dislocation velocity, it becomes a nonlinear condition and possibly even nonlocal, depending upon the constitutive equation for the dislocation velocity. At external inflow boundaries, a zero flux would seem to be appropriate in many circumstances as it is hard to imagine (for this author) an external device that can achieve a non-zero flux at such boundaries. However, at internal surfaces or material interfaces, the boundary condition will need to be specified.

## *8. Final Remarks*

It has been shown in this paper and [I] that a nominally well-posed theory of crystal plasticity can be developed that encompasses many important features of dislocation mechanics. In particular, long-range stresses of dislocation distributions [I] and important physical behaviors related to dislocation density evolution are analytically shown to be natural outcomes of the theory. Because of its field-theoretic nature, it is amenable to computational approaches like the finite element method. In terms of computational complexity for problems involving dense dislocation distributions typical of cold worked metals, an implementation of the theory would stand somewhere between conventional slip-based crystal plasticity and a full discrete dislocation plasticity approach.

The structure of the theory appears to be fairly rigorous on both physical and applied mathematical grounds. Two exceptions that may be noted are the choices of the inner-products used to define the orthogonal projections in the spaces of square integrable matrix fields and $\mathbb{R}^{6K}$, respectively. While sound on mathematical grounds, why they should be appropriate choices on physical grounds may be questioned. To some extent, the analysis of driving forces vindicates the choice of the $L_2$ inner product for square-integrable matrix fields. As for the other choice (Section 4), no compelling physical justification for the choice is apparent to the author. Mathematically, it is perhaps the simplest and most robust idea to achieve the desired physical result – a redistribution of the 'normal' slip system dislocation density increments into 'in-plane' components.

Another physical (and somewhat philosophical) shortcoming, presumably characteristic of any genuinely nonlocal theory, is one related to the analysis of parts of a body. As it stands, the definition of the body is intimately intertwined with the constitutive structure. Consequently, if a subpart of the body were to be analyzed by prescribing a boundary condition history and initial condition on the subpart resulting from an analysis corresponding to the whole body, it seems that a knowledge of the 'whole' body would still be required for the analysis of the subpart to be conducted.

The theory of crystal plasticity presented here suggests a simpler approximation if one is willing to sacrifice detailed crystal structure in the inelastic response. Such a theory would be characterized by only one dislocation density tensor and hence by the following equations:

$$curl\, \boldsymbol{U}^p = -\boldsymbol{\alpha} \tag{78}$$

$$\boldsymbol{U}^p_{/\!/} = \tilde{\boldsymbol{U}}^p_{/\!/} \tag{79}$$

$$\boldsymbol{T} = \boldsymbol{C}\left(\boldsymbol{\varepsilon} - \boldsymbol{\varepsilon}^p\right) \; ; \; \boldsymbol{\varepsilon} := \frac{1}{2}\left(\boldsymbol{U} + \boldsymbol{U}^T\right) \; ; \; \boldsymbol{\varepsilon}^p := \frac{1}{2}\left(\boldsymbol{U}^p + \boldsymbol{U}^{pT}\right) \tag{80}$$

$$div\, \boldsymbol{T} = \boldsymbol{0} \tag{81}$$

$$\dot{\boldsymbol{\alpha}} = -curl\left(\boldsymbol{\alpha} \times \boldsymbol{V}\right) + \boldsymbol{s} \tag{82}$$

$$\mathring{\dot{U}}^p = \boldsymbol{\alpha} \times \boldsymbol{V} \,. \tag{83}$$

The boundary condition (64) would be applicable in this case. The driving forces for the dissipative mechanisms in (78)-(83) may be derived using similar reasoning as for the crystal plasticity case. To ensure deviatoric plasticity when expected, the stress fields appearing in the driving force expressions for the dislocation velocity and nucleation rate may be taken as the deviatoric stress field. A finite element implementation of (78)-(83) is being pursued at the current time.

Finally, the simplified theory as well as the crystal plasticity theory derived herein are nonlocal, but contain no higher-order stresses and require no higher-order boundary conditions that are hard to justify physically (Fleck and Hutchinson, 1997; Shu and Fleck, 1999), neither do they involve boundary conditions on microforces or slips as in Gurtin (2001). The theories presented herein incorporate the elastic theory of dislocation distributions and internal stress exactly – in doing so, the nature of the nonlocal dependence of the stress and the free energy on the dislocation density is found to be through the elastic strain. It is the elastic distortion that is found to be genuinely (integral) nonlocal in the dislocation density field and under the standard *local* dependence of the stress and free energy on the elastic strain, important physical behaviors related to internal stress and motion of dislocation distributions is predicted. This is in contrast to other continuum proposals incorporating continuum measures of dislocation density, e.g. Naghdi and Srinivasa (1993 a, b), Le and Stumpf (1996), Menzel and Steinmann (2000), Gurtin (2001), where the free energy is itself modified by an explicit dependence on the local value of the dislocation density. Such theories have not been shown to reproduce the results of the elastic theory of continuously distributed dislocations. Also, in the theory presented in this paper, the tensorial plastic strain rate is inferred from the dislocation density, in general nonlocally and through a non-invertible relationship, with an additional dependence on the stress dependent dislocation velocity field, as would be expected from the physics of dislocation motion – slip rate is an outcome, not a constitutive specification. In conventional crystal plasticity as well as other gradient theories of continuum plasticity the evolution of slip (plastic deformation) is constitutively specified and the (geometrically necessary) dislocation density is inferred.